\documentclass[onecolumn]{IEEEtran}
\usepackage[latin1]{inputenc}
\usepackage[english]{babel}
\usepackage[margin=1in]{geometry}

\usepackage{blindtext}
\usepackage{amssymb}
\usepackage{hyperref}
\hypersetup{
    colorlinks=true,
    linkcolor=blue,
    filecolor=magenta,      
    urlcolor=cyan,
}
 
\usepackage{multirow}
\usepackage{makecell}
\usepackage{amsthm}
\usepackage[normalem]{ulem}
\usepackage{amsmath}
\usepackage{booktabs}
\usepackage{array}
\usepackage{cite}
\usepackage{soul}
\usepackage{mathtools}
\usepackage{bm}

\newtheorem{theorem}{Theorem}[section]
\newtheorem{corollary}{Corollary}[section]
\newtheorem{lemma}[theorem]{Lemma}
\newtheorem{proposition}[theorem]{Proposition}

\newtheorem{problem}{Problem}

\theoremstyle{definition}
\newtheorem{definition}[theorem]{Definition}

\newcommand{\cB}{\mathcal{B}}
\newcommand{\cC}{\mathcal{C}}

\newcommand{\cM}{\mathcal{M}}
\newcommand{\cN}{\mathcal{N}}
\newcommand{\cO}{\mathcal{O}}

\newcommand{\bcC}{\bar{\mathcal{C}}}

\DeclarePairedDelimiter\parenv{\lparen}{\rparen}

\newcommand{\abs}[1]{\left|#1\right|}
\newcommand{\ceilenv}[1]{\left\lceil #1 \right\rceil}

\newcommand{\bal}[1]{\begin{align}\label{#1}}
\newcommand{\eal}{\end{align}}

\renewcommand{\le}{\leqslant}
\renewcommand{\leq}{\leqslant}
\renewcommand{\ge}{\geqslant}
\renewcommand{\geq}{\geqslant}

\newcommand{\N}{\mathbb{N}}
\newcommand{\R}{\mathbb{R}}
\newcommand{\Z}{\mathbb{Z}}

\newcommand{\kp}{{k_+}}
\newcommand{\km}{{k_-}}
\newcommand{\ve}{\mathbf{e}}

\newcommand{\vv}{\mathbf{v}}

\newcommand{\vx}{\mathbf{x}}
\newcommand{\vy}{\mathbf{y}}

\newcommand{\vc}{\mathbf{c}}

\newcommand{\vp}{\mathbf{p}}

\newcommand{\Zero}{{\mathbf{0}}}

\DeclareMathOperator{\wt}{wt}

\DeclareMathOperator{\supp}{supp}
\newcommand{\eqdef}{\triangleq}

\NewDocumentCommand{\setargsaux}{mm}
{\IfNoValueTF{#2}{#1} {#1\, : \,\mathopen{}#2}}%{#1\:;\:#2}

\title{Elias-type Bounds for Codes in the Symmetric Limited-Magnitude Error Channel}
\author{Zhihao~Guan and Hengjia~Wei
\thanks{This work was supported in part by the National Natural Science Foundation of China under Grant 12371523.}
\thanks{Z. Guan (guanzh@stu.xjtu.edu.cn) and H. Wei (hjwei05@gmail.com) are with the School of Mathematics and Statistics, Xi'an Jiaotong University, Xi'an 710049, China.}
}

\date{}

\begin{document}
\maketitle

% 
% \ead{guanzh@stu.xjtu.edu.cn}

% \author[1]{Hengjia Wei\corref{cor1}}
% \ead{hjwei05@gmail.com}

% \cortext[cor1]{Corresponding author}
% \address[1]{ School of Mathematics and Statistics, Xi'an Jiaotong University, Xi'an 710049, China}

\begin{abstract}
    We study perfect error-correcting codes in $\mathbb{Z}^n$ for the symmetric limited-magnitude error channel, where at most $e$ coordinates of an integer vector may be altered by a value whose magnitude is at most $s$. Geometrically, such codes correspond to tilings of $\mathbb{Z}^n$ by the symmetric limited-magnitude error ball $\mathcal{B}(n,e,s,s)$.  Given $n$ and $s$, we adapt the geometric ideas underlying the Elias bound for the Hamming metric to the distance $d_s$ tailed for this channel, and derive new necessary conditions on $e$ for the existence of perfect codes / tilings, without assuming any lattice structure. Our main results identify two distinct regimes depending on the error magnitude. For small error magnitudes ($s \in \{1, 2\}$), we prove that if the number of correctable errors does not exceed a certain fraction of $n$, then it is asymptotically bounded by $e = \mathcal{O}(\sqrt{n \log n})$. In contrast, for larger magnitudes ($s \geq 3$), we establish a significantly sharper bound of $e < \sqrt{12.36n}$, which holds without any restriction on $e$ being below a given fraction of $n$. Finally, by extending our method to non-perfect codes, we derive an upper bound on packing density, showing that for codes correcting a linear or $\Omega(\sqrt{n})$ number of errors, the density is bounded by a factor inversely proportional to the error magnitude $s$.
\end{abstract}

\section{Introduction}
In many modern data-storage systems, information is encoded as an integer vector $x\in \Z^n$. This representation is fundamental to applications such as multilevel flash memories \cite{cassuto2010codes,schwartz2011quasi} and DNA-based data storage \cite{Jainetal2020ISIT,LeeKalGoeBolChur19,WeiSch21}. These channels are frequently affected by noise in the form of \textit{limited-magnitude errors}, where the values of vector entries are altered by a bounded amount. Developing efficient error-correcting codes for these channels is therefore of significant practical and theoretical importance.

The efficiency of an error-correcting code is quantified by its \textit{rate}. A higher rate indicates greater spatial efficiency, but it is fundamentally constrained by the sphere-packing bound. A code that achieves this bound is called a \textit{perfect code}, representing the most efficient structure possible. This paper investigates the existence of such codes by studying their geometric properties.

The \textit{limited-magnitude error model} was first developed by \cite{cassuto2010codes,schwartz2011quasi}. In this model, at most $e$ entries of the transmitted vector can be altered, with each entry changing by at most $\kp$ in the positive direction or by at most $\km$ in the negative direction. For integers $n \geq e \geq 1$ and $\kp \geq \km \geq 0$, the \emph{$(n,e,\kp,\km) $-error-ball} is defined as
$$
\cB(n,e,\kp,\km) \triangleq \left\{ \vx = (x_1, x_2, \dots, x_n)\in \Z^n:-\km \leq x_i \leq \kp \text{ and } \wt(\vx) \leq e \right\},
$$
where $\wt(\vx)$ denotes the Hamming weight of $\vx$. It is already checked in \cite{WeiSch22EJC} that an error-correcting code in this setting is equivalent to a packing of $\Z^n$ by the error-balls $\cB(n,e,\kp,\km)$ and a perfect code is equivalent to a tiling of $\Z^n$ by $\cB(n,e,\kp,\km)$. 

The study of perfect codes and tilings in the limited-magnitude error model has a rich history. %Much of the foundational research has been on the general asymmetric model. 
For the single-error case ($e=1$), the error balls $\cB(n,1,\kp,\kp)$ and $\cB(n,1,\kp,0)$, commonly reffered to  as crosses and semi-crosses, have been extensively studied \cite{hamaker1984combinatorial,hickerson1986abelian,klove2011some,stein1984packings,stein1994algebra}. These researches were later extended to quasi-crosses, $\cB(n,1,\kp,\km)$, in \cite{schwartz2011quasi}, sparking a surge of activity on the subject \cite{schwartz2014non,yari2013some,ye2019some,zhang2016new,zhang2017nonexistence,zhang2017splitter}. 
More recently, significant progress has been made for the case of two errors ($e=2$), where necessray and sufficient conditions of the existence of tilings have been fully resolved for several small error magnitudes \cite{WeiSch22EJC,zhang2023lattice,guan2025lattice,leung2025latticetilingsmathbbznlimited}.

However, for the case $e>2$, the amount of prior work is relatively limited. Tilings of $\mathbb{Z}^n$ by $\mathcal{B}(n,n-1,k,0)$ have been studied in \cite{buzaglo2012tilings,klove2011some,stein1990notched}. The works \cite{wei2021lattice,WeiSch22EJC} initiated a systematic study of the tiling and packing problems for the general range $2 \le e \le n-1$\footnote{When $e=n$, the ball $\mathcal{B}(n,n,\kp,\km)$ becomes a hypercube, which tiles $\Z^n$ via $((\kp+\km+1)\Z)^n$.}. In these papers, a lattice-tiling construction based on perfect codes in the Hamming metric was provided, together with several constructions of lattice packings that achieve asymptotically nonvanishing  densities. Non-existence results, mostly for lattice tilings, were derived in \cite{WeiSch22EJC}, and these results primarily focus on the regime $e \ge n/2$, with the main findings summarized as follows. 
\begin{theorem}[\cite{WeiSch22EJC}]\label{thm:preresult}
Let  $2\leq  e < n\leq 2e$, and $\kp\geq \km\geq 0$ not both $0$. If $\cB(n,e,\kp,\km)$ lattice tiles $\Z^n$, then one of the following holds:
\begin{enumerate}
\item $\km=0$ and one of the following holds:
 \begin{enumerate}
  \item $e=n-1$(such tilings  have been constructed in \cite{BuzEtz12,stein1990notched});
  \item $(2n-2)/3 \leq e \leq n-3$ and $\kp=1$\footnote{Recall that the entire case of $e=n-2$ has been excluded in  \cite{BuzEtz12}.};
  \item $n/2\leq e<(2n-2)/3$;
 \end{enumerate}
\item $\kp=\km$ and one of the following holds:
 \begin{enumerate}
  \item $(4n-2)/5 \leq e \leq n-1$ and $\kp=\km=1$;
  \item $n/2 \leq e<(4n-2)/5$ and $\sum_{i=1}^e\binom{n}{i} (2\kp)^{i-1} \geq (\kp+1)^{e}$.
 \end{enumerate}
\end{enumerate}
\end{theorem}

This paper contributes to this area by focusing on the important sub-problem of the symmetric limited-magnitude error model, where $\kp=\km=s$. By adapting the geometric principles of the Elias bound, we establish new constraints on the existence of perfect codes in this symmetric setting. Our main results can be stated as follows:

\begin{theorem}
    For small error magnitudes $s \in \{1, 2\}$, if a tiling of $\Z^n$ by  $\cB(n,e,s,s)$ exists, the number of correctable errors $e$ satisfies the following constraints:
    \begin{itemize}
        \item For $s=1$, if $n\in [3,1590]$, then either $e\geq \frac{1}{2}n-\lceil \log_2 n\rceil$ or $e<\sqrt{2n\log_2 n}$; if $n\geq 1591$, then either $e>\frac{3}{5}n$ or $e<\sqrt{9.92 n\log_2 n}$.
        \item For $s=2$, if $n\in [3,1200]$, then either $\frac{3}{4}n-\left\lceil \log_{4/3} \left(\frac{3}{2}n\right)\right\rceil\leq e < \frac{4n-2}{5}$ or $e<\sqrt{\frac{12}{5}n\log_{4/3} \left(\frac{3}{2}n\right)}$; if $n\geq 1201$, then either $\frac{11}{15}n<e<\frac{4n-2}{5}$ or $e<\sqrt{8.80 n\log_2 n}$.
    \end{itemize}
     
\end{theorem}

\begin{theorem}
    For  $s \ge 3$ and $n \geq 61$, if a tiling of $\Z^n$ by $\cB(n,e,s,s)$ exists, then the number of correctable errors $e$ satisfies:
    \begin{itemize}
        \item For  $s \ge 4$, we have $e < \sqrt{12.36n}$.
        \item For $s=3$,  either $e < \sqrt{12.36n}$ or $\frac{2}{3}(n-1) < e < \frac{4n-2}{5}$ with $n\leq 1347$.
    \end{itemize}
\end{theorem}

These results substantially restrict the range of parameters for which perfect codes can exist in this model. The core geometric insight of our method lies in analyzing the average number of codewords within a larger ball, which is an adaptation of the Elias bound, rather than focusing solely on the disjoint error balls centered at each codeword. This approach naturally extends to packings and provides an upper bound on the packing density.

\begin{theorem}
     Let $\delta$ be the packing density of a code correcting $e$ symmetric limited-magnitude errors. If $(e+1)^2 > 2n$, then
     $$\delta \leq \frac{1}{\frac{(e+1)^2}{n}- 2} \cdot \frac{e(e+1)}{s(n-e)}.$$
     Consequently, the following asymptotic bounds hold as $n \to \infty$:
     \begin{enumerate}
         \item If $e = a\sqrt{n}$ with $a>\sqrt{2}$, the packing density is bounded by
         $$\delta\leq \frac{a^2}{s(a^2-2)}+\cO(n^{-1/2}).$$
         \item If $e = an$ with $0 < a < 1$, the packing density is bounded by
         $$\delta\leq \frac{1}{s(1-a)}+\cO(n^{-1}).$$
     \end{enumerate}   
\end{theorem}

\section{Preliminary}
Let $\Z$ denote the ring of integers and $\N$ denote the set of natural numbers. For integers $a \leq b$ we define $[a,b] \triangleq \{a,a +1,...,b\}$ and $[a,b]^* \triangleq [a,b] \backslash \{0\}$. Vectors will be written using bold lower-case letters. For a vector $\vv =(v_1,...,v_n)$, we shall conveniently use $\vv[i]$ to denote its $i$-th entry, namely, $\vv[i] \triangleq v_i$. The support of a vector $\vv$ is the set of its nonzero positions, $\supp(\vv)=\{i:\vv[i]\neq 0\}$, and its Hamming weight is the size of its support.

Let
$$Q=\{(x_1,\ldots,x_n) | 0 \leq x_i<1, x_i \in \R\}$$
denote the semi-open unit hypercube centered at the origin. An integer vector $\vv\in \Z^n$ can be geometrically associated with the translated hypercube $\vv+Q$. Consequently, any discrete set $\cB\subseteq \Z^n$ can be viewed as a cluster of unit hypercubes in $\R^n$.

We say that $\cB \subseteq \Z^n$ \textit{packs} $\Z^n$ by $T \subseteq \Z^n$ if the translates of $\cB$ by elements of $T$ are pairwise disjoint, that is, for all distinct $\vv,\vv' \in T$, 
$$(\vv+\cB)\cap(\vv' +\cB)=\varnothing.$$
We say $\cB$ \textit{covers} $\Z^n$ by $T$ if
$$\bigcup_{\vv\in T}(\vv +\cB)=\Z^n.$$
If $\cB$ both packs and covers $\Z^n$ by $T$, then we say $\cB$ \textit{tiles} $\Z^n$ by $T$. The \textit{packing density} is defined as the asymptotic ratio of the volume covered by the balls to the total volume
$$\delta \eqdef \lim_{L\to +\infty}\frac{|[-L,L]^n\cap T|\cdot|\cB|}{|[-L,L]^n|}.$$
A standard sphere-packing argument implies that $\delta\leq 1$.  

Throughout this paper, we focus on the \textit{symmetric limited-magnitude error model}, where at most $e$ entries of a transmitted vector are altered, and each alteration has a magnitude of at most $s$, implying $\kp=\km=s$. The error ball, denoted as $V(n,e,s)\triangleq \cB(n,e,s,s)$, is defined as the set of all vectors in $\Z^n$ with Hamming weight at most $e$ and entries in $[-s,s]$. An \textit{error-correcting code} (or simply code) $\cC \subseteq \Z^n$ is a set of vectors, called codewords, such that the balls of radius $e$ (representing error-correcting capability) centered at each codeword are pairwise disjoint. This is equivalent to a packing of $\Z^n$ by the error ball $V(n,e,s)$. A \textit{perfect code} is a code meeting the sphere-packing bound; that is, the packing formed by the error balls is also a tiling of $\Z^n$.

Analogous to the Hamming metric, the following distance is used in this error model to characterize the error-correcting capability of a code. 

\begin{definition}
    For two vectors $\mathbf{x},\mathbf{y}\in \mathbb{Z}^n$, we define the distance $d_s(\vx,\vy)$ as:
    $$d_s(\vx,\vy)\triangleq\begin{cases}
        N_s(\mathbf{x},\mathbf{y})+2 M_s(\mathbf{x},\mathbf{y}) \quad &\text{if } \max_i \{|\mathbf{x}[i]-\mathbf{y}[i]| \} \leq 2s,\\
        2n+1 \quad &\text{if } \max_i \{|\mathbf{x}[i]-\mathbf{y}[i]| \}>2s,
    \end{cases}$$
    where
    \begin{align*}
        N_s(\mathbf{x},\mathbf{y}) &\triangleq |\{i|1 \leq |\mathbf{x}[i]-\mathbf{y}[i]|\leq s\}|, \\
        M_s(\mathbf{x},\mathbf{y}) &\triangleq |\{i|s+1 \leq |\mathbf{x}[i]-\mathbf{y}[i]|\leq 2s\}|.
    \end{align*}
\end{definition}

\begin{proposition}\label{prop:distance}
A code $\cC \subset \Z^n$ can correct $e$ $(s,s)$-limited-magnitude errors if and only if $d_s(\vx,\vy) \geq 2e+1$ for all distinct $\vx,\vy\in \cC$.
\end{proposition}
This result parallels Proposition 11 in \cite{WeiSch22} and is adapted for our specific distance definition. For completeness, a proof is provided in Appendix~\ref{appendix:prop}.

\section{Necessary Conditions on the Existence of Perfect Codes}

In this section, we derive necessary conditions for the existence of perfect codes. Our approach adapts the geometric ideas underlying the Elias bound to the limited-magnitude distance metric $d_s$. The overall strategy can be summarized as follows:
\begin{enumerate}
\item \textbf{Averaging:} First, we apply an averaging argument to show that there exists a ball of radius $t>e$ containing a certain number of codewords, which we denote by $K$.
\item \textbf{Quadratic Programming:} Next, we establish an upper bound on the average pairwise distance among these $K$ codewords by formulating the problem as a constrained quadratic program.
\item \textbf{Distance Contradiction:} Finally, we compare this upper bound on the average distance with the minimum required distance of $2e+1$ for an error-correcting code. This comparison yields an inequality that constrains the possible values of $e$.
\end{enumerate}

The main difficulty lies in the second step, where we solve the constrained quadratic program separately for small $s$ ($s \le 3$) and for larger $s$.

\subsection{Codewords in a Larger Ball}

The first step in our approach is to prove that there exists  a ball of radius $t>e$ which contains  at least $\left\lceil|V(n,t,s)|/|V(n,e,s)|\right\rceil$ codewords.

\begin{lemma}\label{lemma:number_of_codewords}
    Let $\cC\subset \Z^n$ be a perfect code correcting $e$ $(s,s)$-limited-magnitude errors. Then for every $t>e$ there exists a vector $\vx \in \Z^n$ such that the set $V(n,t,s)+\vx$  contains at least
    $$\left\lceil\frac{|V(n,t,s)|}{|V(n,e,s)|}\right\rceil$$
    codewords of $\cC$.
\end{lemma}
\begin{IEEEproof}
    Let $L$ be a sufficiently large integer. Denote the $n$ dimensional hypercube $B_L=[-L,L]^n$. Let $M_L\eqdef |\cC\cap B_L|$ be the number of codewords within the box. Since $\cC$ indicates a tiling of $\Z^n$, the translates of the error ball $V(n,e,s)$ by the codewords form a partition of the space. The error balls centered at codewords in $B_L$ cover at least the inner box $B_{L-s}$ and are contained within the outer box $B_{L+s}$. Therefore, the total volume satisfies
    $$(2L-2s+1)^n \leq M_L \cdot |V(n,e,s)| \leq (2L+2s+1)^n.$$
    Consequently, the number of codewords grows asymptotically as
    \begin{equation}\label{eq:M_L}
        M_L=\frac{(2L)^n+\cO(L^{n-1})}{|V(n,e,s)|}.
    \end{equation}
    Next, we employ an averaging argument by considering the total number of codewords contained in all balls $V(n,t,s)$ centered at points $\mathbf{x} \in B_L$ 
    $$S=\sum_{\vx\in B_L}|(\vx+V(n,t,s))\cap\cC|.$$
    By changing the order of summation, this is equivalent to summing the number of balls centered in $B_L$ that cover each codeword $\vc \in \cC$
    $$S=\sum_{\vc\in\cC}|B_L\cap (\vc-V(n,t,s))|.$$
    For any codeword $\vc$ located deep inside the box (specifically, $\vc\in B_{L-t}$), the entire shifted ball $\vc-V(n,t,s)$ is contained within $B_L$. Conversely, for any codeword outside the extended box ($\vc \notin B_{L+t}$), the intersection is empty. Thus, we can bound the sum $S$ as follows
    $$M_{L-t}\cdot |V(n,t,s)| \leq S \leq M_{L+t}\cdot |V(n,t,s)|.$$
    From \eqref{eq:M_L}, both $M_{L-t}$ and $M_{L+t}$ are asymptotically $\frac{(2L)^n+\cO(L^{n-1})}{|V(n,e,s)|}$. Hence, the average number of codewords per ball centered in $B_L$ is
    \begin{equation*}
        \begin{aligned}
            \frac{S}{|B_L|}
            &=\frac{|V(n,t,s)|}{|V(n,e,s)|}\cdot \frac{\left((2L)^n+\cO(L^{n-1})\right)}{(2L+1)^n} \\
            &=\frac{|V(n,t,s)|}{|V(n,e,s)|}\cdot(1+\cO(L^{-1})).
        \end{aligned}
    \end{equation*}
    As $L\to\infty$, the average approaches $\frac{|V(n,t,s)|}{|V(n,e,s)|}$. There must exist at least one ball containing $\ceilenv{\frac{|V(n,t,s)|}{|V(n,e,s)|}}$ codewords.
\end{IEEEproof}

The following results provide lower bounds on the ratio $|V(n,e+r,s)|/|V(n,e,s)|$.

\begin{lemma}\label{bound:number_of_codewords_t=e+1}
    For $e<n-1$, we have
    $$\frac{|V(n,e+1,s)|}{|V(n,e,s)|}\geq \frac{n-e}{e+1}\cdot 2s.$$ 
\end{lemma}

\begin{IEEEproof} We prove it by induction. 
    For $e=1$, we have that $$\frac{|V(n,2,s)|}{|V(n,1,s)|}=\frac{1+2ns+2n(n-1)s^2}{1+2ns}\geq \frac{n-1}{2}\cdot 2s.$$
    For $e<n-1$, we assume  the induction hypothesis: $$\frac{|V(n,e,s)|}{|V(n,e-1,s)|}\geq \frac{n-e+1}{e}\cdot 2s.$$
    Then $$\sum_{i=0}^{e}\binom{n}{i}(2s)^i\geq \frac{n-e+1}{e}\cdot 2s \sum_{i=0}^{e-1}\binom{n}{i}(2s)^i>\frac{n-e}{e+1}\cdot 2s\sum_{i=0}^{e-1}\binom{n}{i}(2s)^i.$$
    Since $$\binom{n}{e+1}(2s)^{e+1}=\frac{n-e}{e+1}\cdot 2s\binom{n}{e}(2s)^{e},$$
    we have
    $$\sum_{i=0}^{e+1}\binom{n}{i}(2s)^i\geq \frac{n-e}{e+1}\cdot 2s\sum_{i=0}^{e}\binom{n}{i}(2s)^i.$$
    That is, $$\frac{|V(n,e+1,s)|}{|V(n,e,s)|}\geq \frac{n-e}{e+1}\cdot 2s,$$
    which completes the proof.
\end{IEEEproof}

\begin{corollary}\label{coro:number_of_codewords}
    For $e+r<n-1$, we have
    $$\frac{|V(n,e+r,s)|}{|V(n,e,s)|}\geq \frac{(n-e-r+1)^r}{(e+r)^r}(2s)^r.$$
\end{corollary}
\begin{IEEEproof}
    By using Lemma~\ref{bound:number_of_codewords_t=e+1} repeatedly, we have
    \begin{equation*}
        \begin{aligned}
            \frac{\abs{V(n,e+r,s)}}{\abs{V(n,e,r)}} &\geq \frac{(n-e)(n-e-1)\cdots(n-e-r+1)}{(e+1)(e+2)\cdots(e+r)}\cdot(2s)^r \\
            &\geq \frac{(n-e-r+1)^r}{(e+r)^r}\cdot(2s)^r.
        \end{aligned}
    \end{equation*}
\end{IEEEproof}

\subsection{Cases for Small \texorpdfstring{$s$}{s}}

Since the metric $d_s$ is translation-invariant, we assume without loss of generality that the vector $\vx$ in Lemma~\ref{lemma:number_of_codewords} is $\Zero$. Let $\cC_r\eqdef \cC \cap V(n,e+r,s)$ and $K_r\eqdef \abs{\cC_r}$. Then $K_r\geq \frac{|V(n,e+r,s)|}{|V(n,e,s)|}$. We derive an upper bound on the average pairwise distance among the codewords in $\cC_r$. This bound is obtained by solving a quadratic programming problem where we maximize the total pairwise distance subject to the constraints on the symbol distribution at each coordinate position $i\in [n]$. 

For each $i \in [n]$ and each $x \in [-s,s]$, let $p_{i,x}$ denote the number of codewords in $\cC_r$ whose $i$-th coordinate equals $x$, that is,
$$p_{i,x}\triangleq \left|\{\vc\in \cC_r:\vc[i]=x\}\right|.$$
The symbol distribution at  position $i$ is then given by  
$$\vp_i\triangleq(p_{i,-s},p_{i,-s+1},\ldots,p_{i,s-1},p_{i,s}).$$

By definition of the metric $d_s$, for any two vectors $\mathbf{c}$ and $\mathbf{c}'$ of $\cC_r$ we have 
$$d_s(\vc,\vc')=N_s(\vc,\vc')+2 M_s(\vc,\vc').$$
Hence, the sum of pairwise distances is

\begin{equation*}
    \begin{aligned}
        \sum_{\substack{ \vc\neq \vc' \\ \vc,\vc' \in \bcC_r}} d_s(\vc,\vc')
        =\sum_{\substack{ \vc\neq \vc' \\ \vc,\vc' \in \bcC_r}}N_s(\vc,\vc')+ 2 \sum_{\substack{ \vc\neq \vc' \\ \vc,\vc' \in \bcC_r}} M_s(\vc,\vc')
        =\sum_{i=1}^n\vp_i D_s \vp_i^{T},
    \end{aligned}
\end{equation*}
where $D_s$ is a $(2s+1)\times(2s+1)$ matrix such that
$$(D_s)_{ij}=\begin{cases}
    0 &\text{if } i=j,\\
    1 &\text{if }1\leq |i-j|\leq s, \\
    2 &\text{if } s+1\leq |i-j|\leq 2s.
\end{cases}$$

%$$D_1 = \begin{pmatrix}
%0 & \frac{1}{2} & 1 \\ 
%\frac{1}{2} & 0 & \frac{1}{2}\\
%1 & \frac{1}{2}&0
%\end{pmatrix}, 
%$$
%$$D_2 = \begin{pmatrix} 0 & \frac{1}{2} &  \frac{1}{2} & 1 & 1 %\\ 
%\frac{1}{2} & 0 & \frac{1}{2} & \frac{1}{2} &1 \\ \frac{1}{2} & \frac{1}{2} & 0 & \frac{1}{2} & \frac{1}{2} \\  1 & \frac{1}{2} & \frac{1}{2} & 0 & \frac{1}{2} \\ 1 & 1 & \frac{1}{2} & \frac{1}{2} &0\end{pmatrix}.$$

Let $f_s(\vp_i)=\vp_i D_s \vp_i^T$. It then suffices to solve the following quadratic programming problem:

\begin{problem}\label{problem:1}
    Maximizing  $$\sum_{i=1}^n f_s(\vp_i)$$ subject to $$\sum_{i=1}^n \vp_i(1,\ldots,1,0,1,\ldots,1)^T\leq  K_r(e+r)$$ and 
$$\vp_i (1,1,\ldots,1)^T = K_r \quad \text{for }i\in[n].$$
\end{problem}
The first constraint specifies that the total weight of all codewords is at most $K_r(e+r)$, while the second constraint simply reflects that the set $\cC_r$ contains exactly $K_r$ codewords.

For small values of $s$ (specifically $s=1,2,3$), this quadratic programming problem can be solved precisely. The following lemma provides the maximal value for the distance contribution $f_s(\vp_i)$ for each coordinate.

\begin{lemma}\label{lm:f_s_leq_3}
    Suppose $\vp_i(1,\ldots,1,0,1,\ldots,1)^T=a_i$ and $\vp_i (1,1,\ldots,1)^T=K_r$ for all $i\in [n]$. 
    \begin{enumerate}
        \item If $s=1$, $f_1(\vp_i)$ attains its maximum $2K_r a_i-a_i^2$ when $\vp_i=\parenv*{\frac{a_i}{2}, K_r-a_i,\frac{a_i}{2}}$.
        \item If $s=2$, $f_2(\vp_i)$ attains its maximum $2K_r a_i-\frac{5a_i^2}{6}$ when $\vp_i=\parenv*{\frac{a_i}{3},\frac{a_i}{6}, K_r-a_i,\frac{a_i}{6},\frac{a_i}{3}}$.
        \item If $s=3$, $f_3(\vp_i)$ attains its maximum $2K_r a_i-\frac{3a_i^2}{4}$ when $\vp_i=\parenv*{\frac{a_i}{4},\frac{a_i}{4},0, K_r-a_i,0,\frac{a_i}{4},\frac{a_i}{4}}$.
    \end{enumerate}
\end{lemma}
The maximum values are derived using standard calculus optimization techniques, detailed in Appendix~\ref{appendix:f_s_leq_3}.

By combining the exact solutions in Lemma~\ref{lm:f_s_leq_3} with the lower bound on $K_r$ given in Corollary~\ref{coro:number_of_codewords}, we can now derive our main results for small $s$. In particular, we show that if $e$ is too large, the required minimum distance $2e+1$ exceeds the attainable average distance, leading to a contradiction.

\begin{lemma}\label{thm:s=1}
    Let $\cC$ be a perfect code in $\Z^n$ ($n\geq 3$) that corrects $e$ $(1,1)$-limited-magnitude errors. Then 
\[\textrm{either   }  e\geq \frac{1}{2}n- \lceil\log_2 n\rceil   \textrm{ or } e< \sqrt{2n\log_2 n}. \]
\end{lemma}
\begin{IEEEproof}
    By Lemma~\ref{lm:f_s_leq_3}, we have $$\sum_{i=1}^n f_1(\vp_i) \leq \sum_{i=1}^n\parenv*{ 2K_r a_i-a_i^2},$$
    and since $\sum_{i=1}^n a_i \leq K_r(e+r)$, it follows that
    \begin{equation*}
            \sum_{i=1}^n f_1(\vp_i)\leq 2 K_r^2(e+r)-\sum_{i=1}^n a_i^2 \leq 2K_r^2(e+r)-\frac{K_r^2(e+r)^2}{n}.
    \end{equation*}
    Hence, the average distance among these codewords satisfies
    $$d_{average}=\frac{1}{K_r (K_r-1)} \sum_{i=1}^n f_1(\vp_i) \leq \left(2e+2r-\frac{(e+r)^2}{n}\right)\frac{K_r}{K_r-1}.$$
    
    Take $r=\lceil \log_2 n\rceil$. If $e\geq \frac{1}{2}n-r$, the theorem holds. Otherwise, from Corollary~\ref{coro:number_of_codewords} we have
    $$K_r\geq \frac{2^r(n-e-r+1)^r}{(e+r)^r}>2^r\geq n.$$
    Therefore,
    \begin{equation*}
        \begin{aligned}
            2e+1\leq d_{average} &\leq \left(2e+2r-\frac{(e+r)^2}{n}\right)\frac{K_r}{K_r-1} \\
            &\leq 2e+2r-\frac{(e+r)^2}{n}+\frac{1}{n}\parenv*{2e+2r-\frac{(e+r)^2}{n}} \\
            &<2e+2r-\frac{(e+r)^2}{n}+1.
        \end{aligned}
    \end{equation*}
    This implies   
    $$(e+r)^2<2rn,$$
    and hence $$e< \sqrt{2n\log_2 n}.$$
\end{IEEEproof}

\begin{lemma}\label{thm:s=2}
    Let $\cC$ be a perfect code in $\Z^n$ ($n\geq 3$) that corrects $e$ $(2,2)$-limited-magnitude errors. Then
    \[\textrm{either   }  e\geq \frac{3}{4}n- \left\lceil\log_{4/3}\left(\frac{3}{2} n\right)\right\rceil   \textrm{ or } e<\sqrt{\frac{12}{5}n\log_{4/3} \left(\frac{3}{2}n\right)}. \]
\end{lemma}
\begin{IEEEproof}
    The proof follows the same structure as Lemma 10, utilizing the bound for $f_2(\vp_i)$ from Lemma~\ref{thm:s=1}. Details can be found in  Appendix~\ref{appendix:s=2,3}.
\end{IEEEproof}

\begin{lemma}\label{thm:s=3}
    Let $\cC$ be a perfect code in $\Z^n$ ($n\geq 3$) that corrects $e$ $(3,3)$-limited-magnitude errors. Then
    \[\textrm{either   }  e\geq \frac{5}{6}n- \left\lceil\log_{6/5}\left(\frac{5}{3} n\right)\right\rceil   \textrm{ or } e<\sqrt{\frac{8}{3}n\log_{6/5} \left(\frac{5}{3}n\right)}. \]
\end{lemma}
The proof is analogous to that of Lemma~\ref{thm:s=1}, and the full derivation can be found in Appendix~\ref{appendix:s=2,3}.

The proofs of Lemmas~\ref{thm:s=1}, \ref{thm:s=2} and \ref{thm:s=3} follow an identical structure, differing only in the constants derived from Lemma~\ref{lm:f_s_leq_3} and the choice of the parameter $r$. This underlying method can be generalized to derive asymptotic bounds for the relationship between $e$ and $n$. The following theorem formalizes this generalization and its proof is provided in Appendix~\ref{appendix:asymptotic}

\begin{theorem}\label{thm:asymptotic}
    For any $\epsilon > 0$ and sufficiently large $n$ such that $\left\lceil\frac{\log_2 n}{\log_2 (1+9\epsilon/4)}\right\rceil<\frac{\epsilon n}{2}$, if a perfect code in $\Z^n$ capable of correcting $e$ $(1,1)$-limited-magnitude errors exists, then either $e>(\frac{2}{3}-\epsilon)n$, or $e<\sqrt{\frac{2n\log_2 n}{\log_2 (1+9\epsilon/4)}}$.

    For any $\epsilon > 0$ and sufficiently large $n$ such that $\left\lceil\frac{\log_2 3n/2}{\log_2 (1+25\epsilon/8)}\right\rceil<\frac{\epsilon n}{2}$, if a perfect code in $\Z^n$ capable of correcting $e$ $(2,2)$-limited-magnitude errors exists, then either $e>(\frac{4}{5}-\epsilon)n$, or $e<\sqrt{\frac{12}{5}nr}=\sqrt{\frac{12n\log_2 n}{5\log_2 (1+25\epsilon/8)}}$.
\end{theorem}
Numerical examples for specific values of $\epsilon$ are provided in Table~\ref{table:1} and Table~\ref{table:2}.

\begin{table}[h]
    \centering
    \begin{tabular}{|c|c|c|}
        \hline
        $\epsilon$ & Minimum $n$ & Bound for $e$ \\
        \hline
        $1/10$ & $641$ & $e<\sqrt{6.84 n\log_2 n}$ \\
        \hline
        $1/15$ & $1591$ & $e<\sqrt{9.92 n\log_2 n}$ \\
        \hline
        $1/20$ & $3041$ & $e<\sqrt{13.01 n\log_2 n}$ \\
        \hline
    \end{tabular}
    \caption{Explicit bounds for $e$ when $s=1$.}
    \label{table:1}
\end{table}

\begin{table}[h]
    \centering
    \begin{tabular}{|c|c|c|}
        \hline
        $\epsilon$ & Minimum $n$ & Bound for $e$ \\
        \hline
        $1/10$ & $501$ & $e<\sqrt{6.12 n\log_2 n}$ \\
        \hline
        $1/15$ & $1201$ & $e<\sqrt{8.80 n\log_2 n}$ \\
        \hline
        $1/20$ & $2241$ & $e<\sqrt{11.46 n\log_2 n}$ \\
        \hline
    \end{tabular}
    \caption{Explicit bounds for $e$ when $s=2$.}
    \label{table:2}
\end{table}

\subsection{Cases for Greater \texorpdfstring{$s$}{s}}
In this section, we take $r=1$ and abbreviate $\mathcal{C}_1$ and $K_1$ as $\bcC$ and $K$, respectively.

Although the quadratic programming approach used for small $s$ provides valid bounds, a distinct structural property emerges in $\bcC$ that enables a sharper analysis and yields a stronger bound on $e$. Specifically, we show that for any coordinate $i$ and any nonzero symbol $x \in [-s,s]^*$, there can be at most one codeword in the local neighborhood $\bcC$ with that specific symbol.This powerful constraint, formalized in the following lemma, significantly restricts the symbol distribution beyond the general conditions considered in Problem~\ref{problem:1}.

\begin{lemma}\label{lemma:constraint_for_t=e+1}Let $\cC\subset \Z^n$ be a code that can correct $e$ $(s,s)$-limited-magnitude errors. Then  for any $i\in [n]$ and $x\in [-s,s]^*$, there is at most one codeword $\vc\in \bcC=\mathcal{C} \cap V(n,e+1,s)$ such that $\vc[i]=x$.
\end{lemma}
\begin{IEEEproof}
    Suppose to the contrary that there exist two codewords $\vc,\vc'\in \mathcal{C}\cap V(n,e+1,s)$ such that $\vc[i]=\vc'[i]=x$. Let $\ve_i$ denote the unit vector with a `$1$' in  the $i$th component.  Denote $\ve = \vc -x  \ve_i$ and $\ve'=\vc'-x \ve_i$. Then both $\ve$ and $\ve'$ are elements of  $V(n,e,s)$, and consequently, so are $-\ve$ and $-\ve'$. However, we then have 
    \[\vc-\ve=\vc'-\ve',\]
    which contradicts the assumption that $\mathcal{C}$ can correct $e$ $(s,s)$-limited-magnitude errors.
\end{IEEEproof}

This lemma allows us to formulate a refined quadratic programming problem. Let $p_{i,x}$ and $p_i$ be defined as in the previous section. Lemma~\ref{lemma:constraint_for_t=e+1} implies $p_{i,x} \in \{0, 1\}$ for $x \in [-s,s]^*$. We seek to maximize the sum of pairwise distances among codewords in $\bcC$:

\begin{problem}
    Maximize $$\sum_{i=1}^n f(\mathbf{p}_i)$$
    subject to
    $$\sum_{i=1}^n \mathbf{p}_i \cdot (1,1,\ldots,1,0,1,\ldots,1)^T= K(e+1),$$
    $$\mathbf{p}_i \cdot (1,1,\ldots,1)^T=K \quad \text{for }i\in[n]$$
    and 
    $$p_{i,x}=0 \ or \ 1 \quad \text{for }i\in [n] \text{ and }x\in[-s,s]^*,$$
    where
    $$f(\mathbf{p}_i)=\vp_i D_s \vp_i^T=2\sum_{x=-s}^s p_{i,x}\sum_{y=x-s}^{x-1}p_{i,y}+4\sum_{x=1}^s p_{i,x}\sum_{y=-s}^{x-s-1} p_{i,y}.$$
\end{problem}

\begin{lemma}\label{lemma:Johnson_bound}
    Suppose that there are $K\geq 2$ codewords in $V(n,e+1,s)$. Then the average distance between these codewords is at most
    $$\frac{(2K-1)(e+1)}{K-1}-\frac{K(e+1)^2}{2(K-1)n}.$$
    Moreover, if  $K(e+1) >\left(s+\frac{1}{3}\right) n$, then the average distance between these codewords is at most
    $$\frac{2(K+s)(e+1)}{(K-1)}-\frac{3K(e+1)^2}{2(K-1)n}-\frac{n(s^2+s)}{K(K-1)}.$$
\end{lemma}
\begin{IEEEproof} 
    Let $a_i=\mathbf{p}_i (1,1,\ldots,1,0,1,\ldots,1)^T$, where $i=1,2,\ldots,n$.
    It is worth noting that since $p_{i,x}= 0$ or $1$ for all $x\neq 0$, we have that 
    \begin{equation}
    \label{eq:faltdef}
    \begin{aligned}
    f(\mathbf{p}_i)= &\  2a_i(K-a_i)+2\binom{a_i}{2} \\
    & \ +2|\{\{x,y\}\subset [-s,s]^*:p_{i,x}=p_{i,y}=1,\ |x-y|\geq s+1\}|.
    \end{aligned}
    \end{equation}

    The proof proceeds in three steps. First, we show that $f(\mathbf{p}_i)$ can attain its maximum value for a specific class of symbol distributions $\mathbf{p}_i$. Second, we derive an upper bound on $f(\mathbf{p}_i)$ in terms of $a_i$. Finally, we combine these results to obtain a bound on the total distance.

    We claim that $\mathbf{p}_i$ can take the  form
    $$(1,\ldots,1,0,\ldots,0,K-a_i,0,\ldots,0,1,\ldots,1)$$
    so that $f(\mathbf{p}_i)$ achieves its maximum value.
    If $p_{i,x}=0$ and $p_{i,y}=1$ for some $x>y>0$, define $\mathbf{p}'_i$ such that $p'_{i,x}=1$ and $p'_{i,y}=0$, with the remaining components of $\mathbf{p}'_i$ the same as those of $\mathbf{p}_i$. By Eq.~\eqref{eq:faltdef}, we have that  
    $$f(\mathbf{p}'_i)-f(\mathbf{p}_i)=2\parenv*{\sum_{|z-x|\geq s+1} p_{i,z}-\sum_{|z-y|\geq s+1} p_{i,z}}\geq 0.$$ Similarly, if $p_{i,x}=0$ and $p_{i,y}=1$ for some $x<y<0$, define $\mathbf{p}'_i$ such that $p'_{i,x}=1$ and $p'_{i,y}=0$, with the remaining components of $\mathbf{p}'_i$ the same as those of $\mathbf{p}_i$. Then,  $f(\mathbf{p}'_i)\geq f(\mathbf{p}_i)$. By repeating this step, we prove our claim.
    
    Denote $l_i=|\{x\in [-s,-1]:p_{i,x}=1\}|$ and $r_i=|\{x\in [1,s]:p_{i,x}=1\}|$. According to the claim above, w.l.o.g., we assume that $\mathbf{p}_i$ has the form \[(\overbrace{1,\ldots,1}^{l_i},0,0,\ldots,0,K-a_i,0,\ldots,0,\overbrace{1,\ldots,1}^{r_i}),\] where $l_i+r_i=a_i$.

    If $a_i\leq s+1$, then for every $x\in [-s,-1]$ and $y \in [1,s]$ with $p_{i,x}=p_{i,y}=1$, we have that $\abs{x-y}\geq s+1$. It follows from Eq.~\eqref{eq:faltdef} that  
    \begin{equation*}
        \begin{aligned}
            f(\mathbf{p}_i)
            &=2a_i(K-a_i)+2\binom{a_i}{2}+2l_i r_i \\
            &\leq 2a_i(K-a_i)+a_i(a_i-1)+\frac{a_i^2}{2} \\
            &=(2K-1)a_i-\frac{1}{2}a_i^2.
        \end{aligned}
    \end{equation*}
    If $a_i> s+1$, then there are $x\in [-s,-1]$ and $y \in [1,s]$ with $p_{i,x}=p_{i,y}=1$ such that $\abs{x-y}\leq s$. Note that for the largest $x\in [-s,-1]$ with $p_{i,x}=1$, i.e., $x=-s+l_i-1$, the number of $y \in [1,s]$ such that  $p_{i,y}=1$ and $\abs{x-y}\leq s$ is 
    \[s-(s-l_i+1+s-r_i)=a_i-s-1.\]
    Hence,
    \[\abs{\{ (x,y)\in [-s,-1]\times [1,s] : p_{i,x}=p_{i,y}=1 \textup{ and } \abs{x-y}\leq s  \} } = \frac{(a_i-s)(a_i-s-1)}{2}.\]
    It follows from Eq.~\eqref{eq:faltdef} that
    \begin{equation*}
        \begin{aligned}
            f(\mathbf{p}_i)
            &=2a_i(K-a_i)+2\binom{a_i}{2}+ 2l_ir_i-(a_i-s)(a_i-s-1)\\
            &\leq 2a_i(K-a_i)+2\binom{a_i}{2}+\frac{a_i^2}{2}- (a_i-s)(a_i-s-1)\\
            &=-\frac{3}{2}a_i^2+2(K+s)a_i-s^2-s.
        \end{aligned}
    \end{equation*}
     Note that for any integer $x$, 
     \[-\frac{1}{2}x^2+(2K-1)x \geq -\frac{3}{2}x^2+2(K+s)x-s^2-s\]
    with equality if and only if $x=s$ or $s+1$.
    We denote 
    $$g(x)=
    \begin{cases}
        -\frac{3}{2}x^2+2(K+s)x-s^2-s  &\text{if } x\geq s, \\
        -\frac{1}{2}x^2+(2K-1)x &\text{if } x< s.
    \end{cases}$$
    Then 
    $$ f(\mathbf{p}_i) \leq g(a_i),$$
    and therefore
     \begin{equation*}
        \begin{aligned}
       \sum_{i=1}^n   f(\mathbf{p}_i) \leq  \sum_{i=1}^n g(a_i)&\leq \sum_{i=1}^n \left(-\frac{1}{2}a_i^2+(2K-1)a_i\right) \\
            &\leq n\left(-\frac{1}{2}\left(\frac{K(e+1)}{n}\right)^2+(2K-1)\frac{K(e+1)}{n}\right) \\
            &=-\frac{K^2(e+1)^2}{2n}+(2K-1)K(e+1).
        \end{aligned}
    \end{equation*}
    Consequently, the average distance satisfies
     \begin{equation*}
        \begin{aligned}
             d_{average}\leq \frac{\sum_{i=1}^n f(\mathbf{p}_i)}{K(K-1)}\leq -\frac{K(e+1)^2}{2(K-1)n}+\frac{(2K-1)(e+1)}{K-1}.
        \end{aligned}
    \end{equation*} 
    
    In the following, we consider the case where  $K(e+1)> \left(s+\frac{1}{3}\right)n$ and seek to improve the bound on $d_{average}$. 
    We assume without loss of generality that $a_1,a_2,\ldots,a_l\geq  s$ and $a_{l+1},\ldots,a_n <s$, for some $l\in [1,n]$. Denote $A\eqdef a_1+a_2+\cdots+a_l$ and $B \eqdef a_{l+1}+\cdots +a_n$. Then $A+B=K(e+1)$.  
    For $1\leq l\leq n-1$, since $g(x)$ is a piecewise concave function, by Jensen's inequality, we have
    \begin{equation*}
        \begin{aligned}
            \sum_{i=1}^n g(a_i)&\leq l g\left(\frac{a_1+\cdots+a_l}{l}\right)+(n-l)g\left(\frac{a_{l+1}+\cdots+a_n}{n-l}\right) \\
            &=l\left(-\frac{3}{2}\left(\frac{A}{l}\right)^2+2(K+s)\frac{A}{l}-s^2-s\right) \\
            &+(n-l)\left(-\frac{1}{2}\left(\frac{B}{n-l}\right)^2+(2K-1)\frac{B}{n-l}\right) \\
            &=-\frac{3}{2}\frac{A^2}{l}+2(K+s)A-(s^2+s)l \\
            &-\frac{(K(e+1)-A)^2}{2(n-l)}+(2K-1)(K(e+1)-A) \\
            &=h(A,l),
        \end{aligned}
    \end{equation*} 
    where
    \begin{equation*}
        \begin{aligned}
            h(A,l)
            &\eqdef \left(-\frac{3}{2l}-\frac{1}{2(n-l)}\right)A^2+\left(\frac{K(e+1)}{(n-l)}+1+2s\right)A-(s^2+s)l \\
            &-\frac{1}{2(n-l)}K^2(e+1)^2+(2K-1)K(e+1).
        \end{aligned}
    \end{equation*}
    For $l=n$, we have
    \begin{equation*}
        \begin{aligned}
            \sum_{i=1}^n g(a_i)&\leq n g\left(\frac{a_1+\cdots+a_n}{n}\right) \\
            &=n\left(-\frac{3}{2}\left(\frac{K(e+1)}{n}\right)^2+2(K+s)\frac{K(e+1)}{n}-s^2-s\right) \\
            &=h(K(e+1),n),
        \end{aligned}
    \end{equation*}
    where $$h(K(e+1),n)\eqdef -\frac{3}{2}\frac{K^2(e+1)^2}{n}+2(K+s)K(e+1)-(s^2+s)n.$$

    Note that $B < s(n-l)$ and $A+B=K(e+1)$. Hence, $A > K(e+1)-s(n-l)$. This bound is better than $A> sl$ as $K(e+1)>sn$. Now, for a fixed $l\in [1,n-1]$, $h(A,l)$ is a quadratic function of $A$ with the axis of symmetry
    $\frac{\frac{K(e+1)}{n-l}+2s+1}{\frac{3}{l}+\frac{1}{n-l}}$.  Compare it with the lower bound $K(e+1)-s(n-l)$ of $A$:
    \begin{equation*}
        \begin{aligned}
            &\ \frac{\frac{K(e+1)}{n-l}+2s+1}{\frac{3}{l}+\frac{1}{n-l}}\leq K(e+1)-s(n-l) \\
            \Longleftrightarrow & \ l\leq 3(K(e+1)-ns).
        \end{aligned}
    \end{equation*}
    The last inequality holds since $l<n<3(K(e+1)-ns)$. Hence, $h(A,l)$ is decreasing with respect to $A$. It follows that 
    \begin{equation*}
        \begin{aligned}
            h(A,l)
            &\leq h(K(e+1)-s(n-l),l) \\
            &= -\frac{3}{2}\frac{(K(e+1)-s(n-l))^2}{l}+2(K+s)(K(e+1)-s(n-l)) \\
            &-(s^2+s)l-\frac{s^2(n-l)}{2}+(2K-1)s(n-l) \\
            &=-\frac{3}{2}(K(e+1)-sn)^2\frac{1}{l}-(2K-s)(K(e+1)-sn)+\frac{s^2}{2}n\\
            &+(2K-1)sn \\
            & \leq -\frac{3}{2}(K(e+1)-sn)^2\frac{1}{n}-(2K-s)(K(e+1)-sn)+\frac{s^2}{2}n\\
            &+(2K-1)sn \\
            & = h(K(e+1),n).
        \end{aligned}
    \end{equation*}
    Therefore, for all $1\leq l\leq n$,  
    $$h(A,l)\leq h(K(e+1),n).$$
Then we have that
\begin{equation*}
        \begin{aligned}
         d_{average}\leq -\frac{3K(e+1)^2}{2(K-1)n}+\frac{2(K+s)(e+1)}{(K-1)}-\frac{n(s^2+s)}{K(K-1)}.
        \end{aligned}
    \end{equation*} 
\end{IEEEproof}

To prove our main result for $s \geq 3$, we require the following bound on $e$, established for general tilings.

\begin{lemma}[\cite{WeiSch22EJC}]\label{lm:preresult}
Let $\cC$ be a perfect code in $\Z^n$ correcting $e$ ($s,s$)-limited-magnitude errors. If $s \ge 2$ and $n \ge 3$, then such a code can only exist if $e < \frac{4n-2}{5}$.
\end{lemma}

\begin{theorem}\label{thm:s_great}
    Let $\cC$ be a perfect code in $\Z^n$ for $n\geq 61$ that corrects $e$ $(s,s)$-limited-magnitude errors where $s\geq 3$.  The error-correcting capability $e$ is bounded as follows:
    \begin{enumerate}
        \item $s\geq 4$, $e<\sqrt{12.36n}$.
        \item $s=3$, either $e<\sqrt{12.36n}$, or $\frac{2}{3}(n-1)<e<\frac{4n-2}{5}$.
    \end{enumerate}
\end{theorem}
\begin{IEEEproof}
    Suppose that there exists a perfect code that can correct $e$ $(s,s)$-limited-magnitude errors.  By Lemma~\ref{lm:preresult}, $e<\frac{4n-2}{5}$. It follows from Lemma~\ref{bound:number_of_codewords_t=e+1} that $$K \geq \frac{n-e}{e+1}\cdot 2s> \frac{2s}{5}>1,$$ so $K\geq 2$.

    If $K(e+1)>\left(s+\frac{1}{3}\right)n$. Applying Lemma \ref{lemma:Johnson_bound}, we obtain
    $$2e+1\leq d_{min}\leq -\frac{3K(e+1)^2}{2(K-1)n}+\frac{2(K+s)(e+1)}{(K-1)}-\frac{n(s^2+s)}{K(K-1)}.$$
    Thus,
    \begin{equation*}
        \begin{aligned}
            0&\leq 1-\frac{3K(e+1)^2}{2(K-1)n}+\frac{2(s+1)(e+1)}{K-1}-\frac{n(s^2+s)}{K(K-1)}\\
            &=1-\frac{1}{K-1}\left(\frac{3K(e+1)^2}{2n}-2(s+1)(e+1)+\frac{n(s^2+s)}{K}\right)\\
            &\leq 1-\frac{1}{K-1}\left(\sqrt{6(s^2+s)}-2(s+1)\right)(e+1)\\
            &\leq 1-\frac{1}{K-1}\left(2\sqrt{2}-\frac{8}{3}\right)s(e+1)
        \end{aligned}
    \end{equation*}
    The second inequality follows from the mean value inequality, while the third inequality is due to the fact that $\sqrt{6\left(1+\frac{1}{s}\right)}-2\left(1+\frac{1}{s}\right)\geq 2\sqrt{2}-\frac{8}{3}$ when $s\geq 3$.
    
    Therefore,
    $$\left(2\sqrt{2}-\frac{8}{3}\right)s(e+1)\leq K-1<\frac{n-e}{e+1}\cdot 2s < \frac{n}{e+1}\cdot 2s,$$
    which leads to
    $$e<\sqrt{\frac{3n}{3\sqrt{2}-4}}-1 < \sqrt{12.36n}-1.$$

   Then we consider the case where $K(e+1)\leq \left(s+\frac{1}{3}\right)n$. 
    Applying Lemma \ref{lemma:Johnson_bound}, we obtain 
    $$2e+1\leq d_{min}\leq d_{average}\leq -\frac{K(e+1)^2}{2(K-1)n}+\frac{(2K-1)(e+1)}{K-1}.$$
    Rearranging this we have 
    \begin{equation}\label{eq:nlbnd-2}
    \left(\frac{(e+1)^2}{n}-2\right)K\leq 2e,
    \end{equation}
    or equivalently,
    \begin{equation}\label{eq:nlbnd}
    n\geq \frac{K(e+1)^2}{2(e+K)}.
    \end{equation}

From Eq.~\eqref{eq:nlbnd-2} and $K \geq \frac{n-e}{e+1}\cdot 2s$, we obtain
    $$\left(\frac{(e+1)^2}{n}-2\right)\frac{n-e}{e+1}\cdot 2s\leq \left(\frac{(e+1)^2}{n}-2\right)K \leq 2e,$$
    which 
    implies 
    $$\frac{(e+1)^2}{n}-2\leq \frac{e(e+1)}{s(n-e)}<\frac{(e+1)^2}{s(n-e-1)}.$$
    Let $\alpha=\frac{e+1}{n}$. We have 
    $$\alpha^2 n-2\leq \frac{\alpha^2 n}{s(1-\alpha)}.$$
    Thus $$n<\frac{2}{\alpha^2 \left(1-\frac{1}{s(1-\alpha)}\right)}.$$

In the following, we estimate $\alpha$.  By Lemma~\ref{thm:preresult}, $e<\frac{4n-2}{5}$, and so $\alpha<\frac{4}{5}$.  
Next, we show that $\frac{n-e}{e+1}\cdot 2s >4$ by considering three cases:
\begin{enumerate}
\item  If $\frac{n-e}{e+1}\cdot 2s\leq 2$, then
        $$e\geq \frac{sn-1}{s+1}.$$
        This contradicts the fact that $e<\frac{2}{3}(n-1)$ when $s=3$  and $e<\frac{4n-2}{5}$ when $s\geq 4$.
\item If $2<\frac{n-e}{e+1}\cdot 2s\leq 3$, then 
        $$\frac{2sn-3}{2s+3}\leq e.$$
        For $s=3$, this contradicts the fact that $e <\frac{2}{3}(n-1)$. For $s\geq 4$, we have 
        \[\frac{8n-3}{11} \leq \frac{2sn-3}{2s+3}\leq e.\]
        Since $K\geq \frac{n-e}{e+1}>2$, necessarily $K\geq 3$. By \eqref{eq:nlbnd}, we have $n\geq \frac{3(e+1)^2}{2(e+3)}\geq \frac{3}{2}e-\frac{3}{2}$. Hence,
        $$\frac{8n-3}{11} \leq e <\frac{2}{3}n+1,$$
        which contradicts $n\geq 21$.
\item If $3<\frac{n-e}{e+1}\cdot 2s\leq 4$, we have  
        $$\frac{3n-2}{5} \leq \frac{sn-2}{s+2}\leq e.$$
        Noting that in this case we have $K\geq 4$,
         by \eqref{eq:nlbnd}, $n\geq \frac{2(e+1)^2}{e+4}\geq 2e-4$. Hence,
        $$\frac{3n-2}{5}\leq e <\frac{1}{2}n+2,$$ which is again a contradiction.
\end{enumerate}
    Therefore, we must have $4<\frac{n-e}{e+1}\cdot 2s$, which implies $e<\frac{sn-2}{s+2}$, and so,  $\alpha<\min\{\frac{s}{s+2},\frac{4}{5}\}$. It follows that $$s(1-\alpha)>s\cdot\max\{\frac{2}{s+2},\frac{1}{5}\}\geq\frac{6}{5}.$$

     On the other hand,
     noting that $K \geq \frac{n-e}{e+1}$, then $\left(s+\frac{1}{3}\right)n\geq 2s(n-e)$. It follows that
    \[\alpha\geq \frac{e}{n} \geq \frac{3s-1}{6s}\geq \frac{4}{9}.\]
    
    Hence,
    $$n<\frac{2}{\alpha^2 \left(1-\frac{1}{s(1-\alpha)}\right)}< 61,$$
    leading to a contradiction.
  
\end{IEEEproof}

In fact, if $n$ is sufficiently large, the additional constraint for the case $s=3$ can be removed.

\begin{corollary}
    If $n\geq 1348$, let $\cC$ be a perfect code in $\Z^n$ that corrects $e$ $(3,3)$-limited-magnitude errors. The error-correcting capability is bounded by 
    $$e<\sqrt{12.36 n}.$$
\end{corollary}
\begin{IEEEproof}
    Combining Lemma~\ref{thm:s=3} and \ref{thm:s_great}, if $n\geq 1348$, we have 
    $$e<\frac{4n-2}{5}<\frac{5}{6}n-\log_{\frac{6}{5}} \frac{5}{3}n-1.$$
    Thus
    $$e<\sqrt{\frac{8}{3}n\log_{\frac{6}{5}} \frac{5}{3}n}<\frac{2}{3}(n-1).$$
\end{IEEEproof}

\section{Extension to Packings and Upper Bounds on Density}

The Elias-type argument developed for perfect codes, which correspond geometrically to tilings of $\Z^n$, can be extended to the more general case of packings. 

%In coding theory, higher packing density generally corresponds to a higher code rate, meaning more codewords per unit volume. However, requiring a code to correct many errors—a large value of $e$—forces the codewords to be far apart. This intuitively limits how densely they can be packed. Our method quantifies this trade-off by establishing an upper bound on the packing density $\delta$ as a function of $n$ and $e$. This result demonstrates that codes with high error-correcting capability must necessarily be sparse, possessing a low packing density.

Similarly to Lemma~\ref{lemma:number_of_codewords}, for a non-perfect code $\cC$, we can also lower bound the number of codewords in a larger ball $V(n,t,s)$.
\begin{lemma}\label{lemma:pakcing_number_of_codewords}
    Let $\cC\subset \Z^n$ be a code correcting $e$ $(s,s)$-limited-magnitude errors with packing density $\delta$. Then for every $t>e$ there exists a vector $\vx \in \Z^n$ such that the set $\vx+V(n,t,s)$  contains at least
    $$\left\lceil\delta\cdot\frac{|V(n,t,s)|}{|V(n,e,s)|}\right\rceil$$
    codewords in $\cC$.
\end{lemma}
\begin{IEEEproof}
    For an integer $L$, denote the $n$ dimensional hypercube $B_L=[-L,L]^n$. Let $M_L\eqdef|\cC\cap B_L|$ be the number of codewords within the box. 
    
    By the definition of packing density, for any $\epsilon>0$ and sufficiently large $L$, we have
    $$\frac{M_L \cdot|V(n,e,s)|}{|B_L|} > \delta-\epsilon. $$
    By an argument analogous to that in Lemma~\ref{lemma:number_of_codewords}, we obtain
    $$\sum_{\vx\in B_L}| (\vx+V(n,t,s))\cap \cC| \geq M_{L-t} \cdot |V(n,t,s)|.$$
    Thus,
    $$\frac{\sum_{\vx\in B_L}| (\vx+V(n,t,s))\cap \cC|}{|B_L|}\geq (\delta-\epsilon)\frac{|V(n,t,s)|}{|V(n,e,s)|}.$$
    Since the maximum must be at least the average, there must exist at least one ball containing 
    $$\left\lceil(\delta-\epsilon)\cdot\frac{|V(n,t,s)|}{|V(n,e,s)|}\right\rceil$$ 
    codewords. Noticing that the ceiling function $\lceil \cdot \rceil$ is left-continuous, for sufficiently small $\epsilon$ we then have
    $$\left\lceil(\delta-\epsilon)\cdot\frac{|V(n,t,s)|}{|V(n,e,s)|}\right\rceil=\left\lceil\delta\cdot\frac{|V(n,t,s)|}{|V(n,e,s)|}\right\rceil.$$
\end{IEEEproof}

\begin{theorem}\label{thm:packing}
    Let $\cC$ be a code in $\Z^n$ that corrects $e$ $(s,s)$-limited-magnitude errors and has a packing density of $\delta$.  If $(e+1)^2>2n$, then its packing density is bounded by 
    $$\delta\leq\frac{1}{\frac{(e+1)^2}{n}-2}\frac{e(e+1)}{s(n-e)}.$$
\end{theorem}
\begin{IEEEproof}
    By Lemma~\ref{bound:number_of_codewords_t=e+1} and \ref{lemma:pakcing_number_of_codewords}, there exists a vector $\vx\in \Z^n$ such that the set $V(n,e+1,s)+\vx $ contains at least
    $$K=\left\lceil\frac{\delta|V(n,e+1,s)|}{|V(n,e,s)|}\right\rceil\geq \delta\left\lceil\frac{|V(n,e+1,s)|}{|V(n,e,s)|}\right\rceil\geq \delta\cdot 2s \cdot\frac{n-e}{e+1}.$$
    codewords. To ensure that $K\geq 2$, it suffices to have $\delta\cdot 2s \cdot\frac{n-e}{e+1}>1$, as $K$ is an integer. Suppose $\delta\leq \frac{e+1}{2s(n-e)}$. Then, since  
    $$\frac{e+1}{2s(n-e)}<\frac{1}{\frac{(e+1)^2}{n}-2}\frac{e(e+1)}{s(n-e)},$$ 
    the bound stated in the theorem is satisfied in this case. Therefore, we may assume $K \geq 2$.
    
    Notably,  Lemma~\ref{lemma:constraint_for_t=e+1} and \ref{lemma:Johnson_bound} also apply to a non-perfect code. The inequality
    $$2e+1\leq d_{min}\leq d_{average}\leq \frac{(2K-1)(e+1)}{K-1}-\frac{K(e+1)^2}{2(K-1)n}$$
    remains valid. Rearranging this inequality gives 
    $$\left(\frac{(e+1)^2}{n}-2\right)K\leq 2.$$
    It follows that
    $$\delta \leq \frac{1}{\frac{(e+1)^2}{n}-2}\frac{e(e+1)}{s(n-e)}.$$
\end{IEEEproof}

\begin{corollary}
    Let $\cC$ be a code in $\Z^n$ correcting $e = a\sqrt{n}$ errors for some constant $a>\sqrt{2}$. For sufficiently large $n$, the packing density $\delta$ of $\cC$ is bounded by
    $$\delta \leq \frac{1}{s} \frac{a^2}{a^2-2} + o(1).$$
\end{corollary}
\begin{IEEEproof}
    Substituting $e=a \sqrt{n}$ into Theorem~\ref{thm:packing}, we obtain
    \begin{equation*}
        \begin{aligned}
            \delta &\leq \frac{1}{a^2-2}\frac{a^2 n+\cO(\sqrt{n})}{sn-\cO(\sqrt{n})} \\
            &=\frac{a^2}{s(a^2-2)}+\cO(n^{-1/2}).
        \end{aligned}
    \end{equation*}
\end{IEEEproof}

\begin{corollary}
    Let $\cC$ be a code in $\Z^n$ correcting $e = a n$ errors for some constant $0<a<1$. For sufficiently large $n$, the packing density $\delta$ of $\cC$ is bounded by
    $$\delta \leq \frac{1}{s(1-a)} + o(1).$$
\end{corollary}
\begin{IEEEproof}
    Substituting $e=a n$ into Theorem~\ref{thm:packing}, we obtain
    \begin{equation*}
        \begin{aligned}
            \delta \leq \frac{1}{a^2 n+\cO(1)}\frac{a^2 n+\cO(1)}{s(1-a)} =\frac{1}{s(1-a)}+\cO(n^{-1}).
        \end{aligned}
    \end{equation*}
\end{IEEEproof}

\section{Conclusion}

This paper investigated the existence of perfect codes capable of correcting symmetric limited-magnitude errors, characterized by the error ball $\cB(n,e,s,s)$. By adapting the geometric principles underlying the Elias bound to the specific distance metric $d_s$, we derived novel necessary conditions on the error-correcting capability $e$.

Our main contributions establish fundamental constraints on the parameters for which such perfect codes can exist. %We demonstrated two distinct regimes based on the error magnitude $s$ and the structural constraints of the code: 
%\begin{itemize}
%    \item Small Error Magnitudes ($s=1, 2$): We demonstrated that the error-correcting capability is asymptotically bounded by $e = \cO(\sqrt{n \log n})$.
%    \item Large Error Magnitudes ($s \ge 3$): For $s \geq 4$, we established a significantly sharper bound of $e < \sqrt{12.36n}$. Notably, our analysis revealed that the case $s=3$ exhibits transitional behavior: while generally subject to the logarithmic bound, it satisfies the sharper condition $e < \sqrt{12.36n}$ for sufficiently large $n$ ($n \geq 1348$).
%\end{itemize}
A significant aspect of our work is that these bounds apply to any perfect code, or tiling, regardless of whether it possesses a lattice structure. This generality stems from the use of a density argument that does not rely on periodicity, broadening the applicability of our results beyond much of the previous work focused primarily on lattice tilings. 

Furthermore, we extended our methodology to non-perfect codes, or packings, deriving an upper bound on the packing density $\delta$. This result quantifies the necessary trade-off between error-correcting capability and code density, proving that codes correcting a high number of errors---specifically when $e = \Omega(\sqrt{n})$---must necessarily be sparse, with density vanishing as $n \to \infty$.

These findings significantly narrow the search space for perfect codes in the symmetric limited-magnitude error model. Future work could focus on tightening these bounds, exploring the asymmetric error model using similar techniques, or investigating specific constructions that might approach these limits. The question of whether perfect codes meeting these bounds exist, particularly in the non-lattice case, remains an intriguing problem.

\bibliographystyle{IEEEtran}
\bibliography{ref}

\appendices

\section{Omitted Proofs}

\subsection{Proof of Proposition~\ref{prop:distance}}\label{appendix:prop}
% \begin{proposition}
% A code $\cC \subset \Z^n$ can correct $t$ $(s,s)$-limited-magnitude errors if and only if $d_s(\vx,\vy) \geq 2t+1$ for all distinct $\vx,\vy\in \cC$.
% \end{proposition}
\begin{IEEEproof}
    If $\cC$ cannot correct $t$ $(s,s)$-limited-magnitude errors, there exist distinct codewords $\vx\neq\vy$ such that $\vx+\ve_1=\vy+\ve_2$ for some $\ve_1,\ve_2,\in V(n,t,s)$. Since $d_s(\vx,\vy)=d_s(\ve_1-\ve_2,\Zero)$, it suffices to show the following two sets are equivalent:
    $$\{\ve\in\Z^n:\ve=\ve_1-\ve_2, \ve_1 \text{ and }\ve_2\in V(n,t,s)\}$$
    and
    $$\{\ve\in\Z^n:d_s(\ve,\Zero)\leq 2t\}.$$

    First, suppose $\ve_1,\ve_2\in V(n,t,s)$. Then they have Hamming weight at most $t$ and entries located in $[-s,s]$. Let 
    $$\cN=\{i:1\leq |\ve_1[i]-\ve_2[i]|\leq s\},$$
    $$\cM=\{i:s+1\leq |\ve_1[i]-\ve_2[i]|\leq 2s\}.$$
    Notice that $\cN\cap\cM=\varnothing$ and $\cM\subseteq\supp(\ve_1)\cap\supp(\ve_2)$. Consider the subsets
    $$\cN_1=\supp(\ve_1)\cap \cN,$$
    $$\cN_2=\supp(\ve_2)\cap \cN.$$
    Then necessarily $\cN=\cN_1\cup \cN_2$, $\cN_1\cup \cM\subseteq \supp(\ve_1)$, $\cN_2\cup \cM\subseteq \supp(\ve_2)$. 
    Thus,
    \begin{equation*}
        \begin{aligned}
            d_s(\ve_1-\ve_2,\Zero)
            &=|\cN|+2|\cM| \\
            &\leq |\cN_1|+|\cN_2|+2|\cM| \\
            &\leq |\supp(\ve_1)|+|\supp(\ve_2)| \\
            &\leq 2t.
        \end{aligned}
    \end{equation*}

    On the other hand, suppose $\ve\in \Z^n$ such that $d_s(\ve,\Zero)\leq 2t$. Let
    $$\cN=\{i:1\leq |\ve[i]|\leq s\},$$
    $$\cM=\{i:s+1\leq |\ve[i]|\leq 2s\}.$$
    For $i\in \cM$, assume $\ve[i]>s$. Let $\ve_1[i]=s$, $\ve_2[i]=\ve[i]-s$. 
    
    Since $|\cN|=d(\ve,\Zero)-2|\cM|\leq 2t-2|\cM|$, we can divide $\cN$ into two disjoint subsets $\cN_1$ and $\cN_2$, such that both of their sizes are at most $t-|\cM|$.
    For $i\in \cN_1$, let $\ve_1[i]=\ve[i]$ and $\ve_2[i]=0$. For $i\in \cN_1$, let $\ve_1[i]=\ve[i]$ and $\ve_2[i]=0$. For $i\in \cN_2$, let $\ve_2[i]=\ve[i]$ and $\ve_1[i]=0$. For $i\notin \cN \cap \cM$, let $\ve_1[i]=\ve_2[i]=0$. Then $\wt(\ve_1)=|\cN_1|+|\cM|\leq t$, $\wt(\ve_2)=|\cN_2|+|\cM|\leq t$.    
\end{IEEEproof}

\subsection{Proof of Lemma~\ref{lm:f_s_leq_3}}\label{appendix:f_s_leq_3}
% \begin{lemma}
%     Suppose $\vp_i(1,\ldots,1,0,1,\ldots,1)^T=a_i$ and $\vp_i (1,1,\ldots,1)^T=K_r$ for all $i\in [n]$. 
%     \begin{enumerate}
%         \item If $s=1$, $f_1(\vp_i)$ attains its maximum $2K_r a_i-a_i^2$ when $\vp_i=\parenv*{\frac{a_i}{2}, K_r-a_i,\frac{a_i}{2}}$.
%         \item If $s=2$, $f_2(\vp_i)$ attains its maximum $2K_r a_i-\frac{5a_i^2}{6}$ when $\vp_i=\parenv*{\frac{a_i}{3},\frac{a_i}{6}, K_r-a_i,\frac{a_i}{6},\frac{a_i}{3}}$.
%         \item If $s=3$, $f_3(\vp_i)$ attains its maximum $2K_r a_i-\frac{3a_i^2}{4}$ when $\vp_i=\parenv*{\frac{a_i}{4},\frac{a_i}{4},0, K_r-a_i,0,\frac{a_i}{4},\frac{a_i}{4}}$.
%     \end{enumerate}
% \end{lemma}
\begin{IEEEproof}
    We first claim that we can assume the optimal distribution $\vp_i$ is symmetric. Let $a_i$ be the fixed weight of the nonzero coordinates, i.e., $\sum_{x \neq 0} p_{i,x} = a_i$, and let $p_{i,0} = K_r - a_i$. 
    We aim to maximize the objective function $f_s(\vp_i) = \vp_i D_s \vp_i^T$. Notice that
    $$f_s(\vp_i)=2\parenv*{\sum_{x\in[-s,s]}p_{i,x}}^2-2\sum_{x\in[-s,s]} p_{i,x}^2 - \sum_{\substack{x,y\in[-s,s] \\ 1\leq |x-y|\leq s}}p_{i,x}p_{i,y}.$$
    Since $\sum_{x\in[-s,s]} \vp_{i,x} = K_r$ is constant, maximizing $f_s(\vp_i)$ is equivalent to minimizing the function
    $$Q_s(\vp_i) = 2 \sum_{x \in [-s,s]^*} p_{i,x}^2 + \sum_{\substack{x,y \in [-s,s]^* \\ 1\leq |x-y| \leq s}} p_{i,x} p_{i,y}.$$
    The function $Q_s(\vp_i)$ is a quadratic form $\vp M_s \vp^T$ restricted to the nonzero indices. By Sylvester's Criterion, the matrix $M_s$ is positive definite for $s \in \{1, 2, 3\}$, implying that $Q_s(\vp_i)$ is strictly convex. Furthermore, the optimization problem is symmetric under reflection ($x \to -x$). Then the objective $Q_s(\vp_i)$ is invariant if we swap $p_{i,x}$ and $p_{i,-x}$, and the constraints ($\sum_{x \in[-s,s]^*} p_{i,x} = a_i$ and $p_{i,x} \geq 0$) are also invariant. Because $Q(\vp_i)$ is strictly convex, it possesses a unique global minimum. Since the problem is symmetric, this unique minimum must lie at the center of symmetry. Therefore, the optimal solution must satisfy $p_{i,x} = p_{i,-x}$.

    For $s=1$, the result is easily obtained.

    For $s=2$, the symmetry condition implies $p_{i,2} = p_{i,-2}$ and $p_{i,1} = p_{i,-1}$. Suppose $\vp_i$ has the form $$\parenv*{\frac{b_i}{2},\frac{a_i-b_i}{2}, K_r-a_i,\frac{a_i-b_i}{2},\frac{b_i}{2}}.$$ Then
    $$f_2(\vp_i)=2a_i K_r+\frac{-3a_i^2+4 a_i b_i-3 b_i^2}{2}\leq 2 K_ra_i-\frac{5}{6}a_i^2,$$
    and equality holds when $b_i=\frac{2 a_i}{3}.$

    For $s=3$, similarly suppose $\vp_i$ has the form $\parenv*{\frac{c_i}{2},\frac{b_i}{2},\frac{a_i-b_i-c_i}{2},K_r-a_i,\frac{a_i-b_i-c_i}{2},\frac{b_i}{2},\frac{c_i}{2}}$. Then
    $$f_3(\vp_i)=2 a_i K_r+\frac{-3a_i^2-b_i^2-3c_i^2+2a_i b_i+4 a_i c_i-2b_i c_i}{2}.$$
    Since $\frac{\partial f_3(\vp_i)}{\partial b_i}=\frac{a_i-b_i-c_i}{2}$, $\frac{\partial f_3(\vp_i)}{\partial c_i}=\frac{2a_i-b_i-3c_i}{2}$, we have $f_3(\vp_i)$ attain its maximum $2K_r a_i-\frac{3}{4}a_i^2$ when $b_i=c_i=\frac{a_i}{2}$. 
\end{IEEEproof}

\subsection{Proofs of Lemmas~\ref{thm:s=2} and \ref{thm:s=3}}\label{appendix:s=2,3}
Both proofs follow the identical logic established in Lemma~\ref{thm:s=1}. In both cases, we maximize the average distance by utilizing the quadratic bounds on $f_s(\vp_i)$ derived in Lemma~\ref{lm:f_s_leq_3}.
% \begin{lemma}
%     Let $\cC$ be a perfect code in $\Z^n$ ($n\geq 3$) that corrects $e$ $(2,2)$-limited-magnitude errors. Then
%     \[\textrm{either   }  e\geq \frac{3}{4}n- \left\lceil\log_{4/3}\left(\frac{3}{2} n\right)\right\rceil   \textrm{ or } e<\sqrt{\frac{12}{5}n\log_{4/3} \left(\frac{3}{2}n\right)}. \]
% \end{lemma}
\begin{IEEEproof}[Proof of Lemma~~\ref{thm:s=2}]
    By Lemma~\ref{lm:f_s_leq_3}, 
    \begin{equation*}
        \begin{aligned}
            \sum_{i=1}^n f_2(\vp_i) &\leq \sum_{i=1}^n \parenv*{ 2K_ra_i-\frac{5a_i^2}{6}}= 2K_r^2(e+r)-\frac{5}{6}\sum_{i=1}^n a_i^2 \\
            &\leq 2K_r^2(e+r)-\frac{5K_r^2(e+r)^2}{6n}.
        \end{aligned}
    \end{equation*}

    The average distance of these codewords
    $$d_{average}\leq \frac{1}{K_r (K_r-1)} \sum_{i=1}^n f_2(\vp_i) \leq \left(2e+2r-\frac{5}{6}\frac{(e+r)^2}{n}\right)\frac{K_r}{K_r-1}.$$

    Take $r=\lceil\log_{\frac{4}{3}} \frac{3}{2}n\rceil$. The condition $e<\frac{3}{4}n-r$ leads to
    $$K_r\geq \frac{4^r(n-e-r+1)^r}{(e+r)^r}>\parenv*{\frac{4}{3}}^r\geq \frac{3}{2}n.$$
    Therefore
    \begin{equation*}
        \begin{aligned}
            2e+1\leq d_{average} &\leq \left(2e+2r-\frac{5}{6}\frac{(e+r)^2}{n}\right)\frac{K_r}{K_r-1} \\
            &\leq 2e+2r-\frac{5}{6}\frac{(e+r)^2}{n}+\frac{2}{3n}\parenv*{2e+2r-\frac{5}{6}\frac{(e+r)^2}{n}} \\
            &<2e+2r-\frac{5}{6}\frac{(e+r)^2}{n}+1.
        \end{aligned}
    \end{equation*}
    Thus 
    $$(e+r)^2<\frac{12}{5}rn,$$
    $$e\geq \sqrt{\frac{12}{5}n\log_{\frac{4}{3}} \frac{3}{2}n}.$$
\end{IEEEproof}

\vspace{1em}

% \begin{lemma}
%     Let $\cC$ be a perfect code in $\Z^n$ ($n\geq 3$) that corrects $e$ $(3,3)$-limited-magnitude errors. Then
%     \[\textrm{either   }  e\geq \frac{5}{6}n- \left\lceil\log_{6/5}\left(\frac{5}{3} n\right)\right\rceil   \textrm{ or } e<\sqrt{\frac{8}{3}n\log_{6/5} \left(\frac{5}{3}n\right)}. \]
% \end{lemma}
\begin{IEEEproof}[Proof of Lemma~\ref{thm:s=3}]
    By Lemma~\ref{lm:f_s_leq_3}, 
    \begin{equation*}
        \begin{aligned}
            \sum_{i=1}^n f_3(\vp_i) &\leq \sum_{i=1}^n \parenv*{ 2K_ra_i-\frac{3 a_i^2}{4}}= 2K_r^2(e+r)-\frac{3}{4}\sum_{i=1}^n a_i^2 \\
            &\leq 2K_r^2(e+r)-\frac{3 K_r^2(e+r)^2}{4 n}.
        \end{aligned}
    \end{equation*}

    The average distance of these codewords
    $$d_{average} \leq \frac{1}{K_r (K_r-1)} \sum_{i=1}^n f_3(\vp_i) \leq \left(2e+2r-\frac{3}{4}\frac{(e+r)^2}{n}\right)\frac{K_r}{K_r-1}.$$

    Take $r=\lceil\log_{\frac{6}{5}} \frac{5}{3}n\rceil$. $e<\frac{5}{6}n-r$ leads to
    $$K_r\geq \frac{6^r(n-e-r+1)^r}{(e+r)^r}>\parenv*{\frac{6}{5}}^r\geq \frac{5}{3}n.$$
    Therefore
    \begin{equation*}
        \begin{aligned}
            2e+1\leq d_{average} &\leq \left(2e+2r-\frac{3}{4}\frac{(e+r)^2}{n}\right)\frac{K_r}{K_r-1} \\
            &\leq 2e+2r-\frac{3}{4}\frac{(e+r)^2}{n}+\frac{3}{5 n}\parenv*{2e+2r-\frac{3}{4}\frac{(e+r)^2}{n}} \\
            &<2e+2r-\frac{3}{4}\frac{(e+r)^2}{n}+1.
        \end{aligned}
    \end{equation*}
    Thus 
    $$(e+r)^2<\frac{8}{3}rn,$$ 
    $$e<\sqrt{\frac{8}{3}n\log_{\frac{6}{5}} \frac{5}{3}n}.$$
\end{IEEEproof}

\subsection{Proof of Theorem~\ref{thm:asymptotic}}\label{appendix:asymptotic}

% \begin{theorem}
%     For any $\epsilon > 0$ and sufficiently large $n$, such that $\left\lceil\frac{\log_2 n}{\log_2 (1+9\epsilon/4)}\right\rceil<\frac{\epsilon n}{2}$, if a perfect code in $\Z^n$ capable of correcting $e$ $(1,1)$-limited-magnitude errors exists, then either $e>(\frac{2}{3}-\epsilon)n$, or $e<\sqrt{\frac{2n\log_2 n}{\log_2 (1+9\epsilon/4)}}$.

%     For any $\epsilon > 0$ and sufficiently large $n$, such that $\left\lceil\frac{\log_2 3n/2}{\log_2 (1+25\epsilon/8)}\right\rceil<\frac{\epsilon n}{2}$, if a perfect code in $\Z^n$ capable of correcting $e$ $(2,2)$-limited-magnitude errors exists, then either $e>(\frac{4}{5}-\epsilon)n$, or $e<\sqrt{\frac{12}{5}nr}=\sqrt{\frac{12n\log_2 n}{5\log_2 (1+25\epsilon/8)}}$.
% \end{theorem}
\begin{IEEEproof}
    \noindent\textbf{Case 1} ($s=1$): If $e \leq (\frac{2}{3} - \epsilon)n$, for sufficiently small $\epsilon>0$, take $$r=\left\lceil\frac{\log_2 n}{\log_2 (1+9\epsilon/4)}\right\rceil.$$
    There exists an integer $N(\epsilon)$ such that for any $n>N(\epsilon)$, the inequality $r<\frac{\epsilon n}{2}$ holds. Consequently, the condition $e \leq (\frac{2}{3} - \epsilon)n$ implies $e + r < (\frac{2}{3} - \frac{\epsilon}{2})n$.
    Substituting this into the lower bound for $K_r$ from Corollary~\ref{coro:number_of_codewords}, we obtain
    \begin{equation*}
        \begin{aligned}
            K_r\geq \frac{2^r(n-e-r+1)^r}{(e+r)^r}> 2^r \parenv*{\frac{\frac{1}{3}+\frac{\epsilon}{2}}{\frac{2}{3}-\frac{\epsilon}{2}}}^r>\parenv*{1+\frac{9}{4}\epsilon}^r\geq n.
        \end{aligned}
    \end{equation*}
    By following the identical calculation steps as in the proof of Lemma~\ref{thm:s=1}, we arrive at the inequality: 
    $$(e+r)^2 < 2rn.$$
    Substituting $r$, we conclude
    $$e<\sqrt{2nr}=\sqrt{\frac{2n\log_2 n}{\log_2 (1+9\epsilon/4)}}.$$

    \noindent\textbf{Case 2} ($s=2$): Similarly, if $e \leq (\frac{4}{5} - \epsilon)n$, for sufficiently small $\epsilon > 0$, let $$r=\left\lceil\frac{\log_2 3n/2}{\log_2 (1+25\epsilon/8)}\right\rceil.$$
    There exists an integer $N(\epsilon)$ such that for any $n>N(\epsilon)$, the inequality $\frac{\epsilon n}{2}>r$ holds. Consequently, the condition $e \leq (\frac{4}{5} - \epsilon)n$ implies $e + r < (\frac{4}{5} - \frac{\epsilon}{2})n$. The lower bound for $K_r$ becomes:
    \begin{equation*}
        \begin{aligned}
            K_r\geq \frac{4^r(n-e-r+1)^r}{(e+r)^r}> 4^r \parenv*{\frac{\frac{1}{5}+\frac{\epsilon}{2}}{\frac{4}{5}-\frac{\epsilon}{2}}}^r>\parenv*{1+\frac{25}{8}\epsilon}^r\geq \frac{3}{2}n.
        \end{aligned}
    \end{equation*}
    By following the identical calculation steps as in the proof of Lemma~\ref{thm:s=2}, we arrive at the inequality: 
    $$(e+r)^2<\frac{12}{5}rn.$$
    Substituting $r$, we conclude
    $$e< \sqrt{\frac{12}{5}nr}=\sqrt{\frac{12n\log_2 n}{5\log_2 (1+25\epsilon/8)}}.$$
\end{IEEEproof}

%For the non-symmetric case, it is challenging to estimate the average distance. However, for the case of $(\kp,\km)=(1,0)$, we still can use the Johnson bound to derive some nonexistence result. Let $$S(n,w,1)\eqdef \set{\vx \in \set{0,1}^n; \wt(\vx)=w}.$$ Let $\cC\subset \Z^n$ be a perfect lattice code that can correct $e$ $(1,0)$-limited-magnitude errors.  Denote $\cC'\eqdef \cC \cap S(n,e+r,1)$. W.l.o.g., we assume that $\abs{\cC'} \geq \frac{\abs{S(n,e+r,1)}}{\abs{\cB(n,e,1,0)}}.$  For any two distinct codewords $\vx,\vy\in \cC'$, we have  $$\abs{\set{i;x_i=1,y_i=0 }} = \abs{\set{i;x_i=0,y_i=1 }}\geq e+1,$$ where the equality holds since $\vx$ and $\vy$ are binary vectors of the same weight and the inequality holds as $d_{1,0} (\vx,\vy) =\max\set{ \abs{\set{i;x_i=1,y_i=0 }}, \abs{\set{i;x_i=0,y_i=1 }} }  \geq e+1$. Hence, $d_H(\vx,\vy)\geq 2e+2$. Applying the Johnson bound for codes in the Hamming metric, we have that $$2e+2 \leq \frac{K}{K-1}\parenv*{2(e+r)-\frac{1}{2}\frac{(e+r)^2}{n} }.$$

\end{document}